\documentclass[12pt]{article}

\usepackage{graphics,graphpap}

\renewcommand{\theequation}{\arabic{equation}}
\newcommand{\bea}{\begin{eqnarray}}
\newcommand{\eea}{\end{eqnarray}}
\newcommand{\beq}{\begin{equation}}
\newcommand{\eeq}{\end{equation}}

\def\msbar{\ifmmode{\overline{\rm MS}} \else{$\overline{\rm MS}$} \fi}
\def\drbar{\ifmmode{\overline{\rm DR}} \else{$\overline{\rm DR}$} \fi}
\def\sf{\ifmmode{\tilde{f}} \else{$\tilde{f}$} \fi}
\def\st{\ifmmode{\tilde{t}} \else{$\tilde{t}$} \fi}
\def\sb{\ifmmode{\tilde{b}} \else{$\tilde{b}$} \fi}
\def\sq{\ifmmode{\tilde{q}} \else{$\tilde{q}$} \fi}
\def\sg{\ifmmode{\tilde{g}} \else{$\tilde{g}$} \fi}
\def\bbar{\ifmmode{\bar{b}} \else{$\bar{b}$} \fi}
\def\tbar{\ifmmode{\bar{t}} \else{$\bar{t}$} \fi}
\def\qbar{\ifmmode{\bar{q}} \else{$\bar{q}$} \fi}
\def\ksla{{k \hspace{-2mm} \slash}}

\newcommand\bvec{\left( \begin{array}{c}}
\newcommand\evec{\end{array}\right)}

\newcommand\bmat{\left( \begin{array}{cc}}
\newcommand\emat{\end{array}\right)}

\renewcommand\d{\delta}

\def\ksla{{k \hspace{-2.2mm} \slash}}

\newcommand{\sW}{\sin\theta_W}

\newcommand\cchp{{\tilde\chi^+}}

\newcommand\cch{{\tilde\chi^0}}

\renewcommand\d{\delta}
\renewcommand\a{\alpha}
\renewcommand\b{\beta}

\def\su{\ifmmode{\tilde{u}} \else{$\tilde{u}$} \fi}
\def\sd{\ifmmode{\tilde{d}} \else{$\tilde{d}$} \fi}

\begin{document}
%------------------------------------------------------------------------

\pagestyle{empty} \vspace*{-1cm}
\begin{flushright}
  HEPHY-PUB 769/03\\
  hep-ph/0304006
\end{flushright}

\vspace*{1.4cm}

\begin{center}
\begin{Large} \bf
Analysis of the chargino and neutralino mass parameters at one--loop level
\end{Large}

\vspace{10mm}

{\large W.~\"Oller, H.~Eberl, W.~Majerotto, C.~Weber}

\vspace{6mm}
\begin{tabular}{l}
 {\it Institut f\"ur Hochenergiephysik der \"Osterreichischen
 Akademie der Wissenschaften,}\\
{\it A--1050 Vienna, Austria}
\end{tabular}

\vspace{20mm}

\begin{abstract}
In the Minimal Supersymmetric Standard Model (MSSM) the masses of
the neutralinos and charginos depend on the gaugino and higgsino mass parameters $M$, $M'$
and $\mu$. If supersymmetry is realized, the extraction of these parameters from future high
energy experiments will be crucial to test the underlying theory.
We present a consistent method how on--shell
parameters can be properly defined at one--loop level and how they
can be determined from precision measurements. In addition, we show how a GUT
relation for the parameters $M$ and $M'$ can be tested at one--loop level. The
numerical analysis is based on a complete one--loop calculation.
The derived analytic formulae are given in the appendix.
\end{abstract}
\end{center}

\vfill

\newpage
\pagestyle{plain} \setcounter{page}{2}

\section{Introduction}
If supersymmetry (SUSY) as the most attractive extension of the
Standard Model is realized at low energies, the next generation of
high energy physics experiments at Tevatron, LHC and a future
$e^+e^-$ linear collider will discover supersymmetric particles.
Particularly at a linear collider, it will be possible to perform
measurements with high precision \cite{tesla, lincol} which allows to test the
underlying SUSY model. For instance, at TESLA \cite{tesla} the
precision of the mass determination of charginos and neutralinos,
the supersymmetric partners of the gauge and Higgs bosons, will be
$\Delta m_{\tilde{\chi}^{\pm,0}} = 0.1 - 1$ GeV. To match this
accuracy it is indispensable to include higher order radiative
corrections.\\
One goal of all analyses based on precise measurements of cross
sections, decay branching ratios, masses of supersymmetric
particles, etc. will be the reconstruction of the fundamental
parameters of the underlying supersymmetric model. In particular, this
is needed for extrapolating the parameters to
the GUT point to check the unification of the supersymmetry
breaking parameters \cite{Tsukamoto, BPZ}. \\
In the Minimal Supersymmetric
Standard Model (MSSM) the chargino and neutralino system depends
on the parameters $M$, $M'$, $\mu$ and $\tan\beta$. $M$ and $M'$ are the
SU(2) and U(1) gauge mass parameter, $\mu$ is the higgsino mass parameter and
$\tan\beta=\frac{v_2}{v_1}$ with $v_{1,2}$ the vacuum
expectation values of the two neutral Higgs doublet fields.
At lowest order, it was shown in \cite{Kalinowski, Kalinowski2}
that these parameters can be extracted from the masses and
production cross sections in $e^+e^-$ collisions with polarized electron beams.
At higher order, this extraction of the parameters is, however,
not trivial. It depends on the definition of the mass matrices (at
higher order) and on the renormalization scheme.
This is the subject of this paper.\\
In the (scale dependent) $\overline{\rm DR}$ scheme the one--loop
corrections to the chargino/neutralino mass matrix were calculated
in \cite{pierce, lahanas}. In \cite{yamada} effective
chargino mixing matrices were introduced, which are independent of
the renormalization scale. For the on--shell renormalization of
the chargino and neutralino system which we adopt here, two methods were
proposed \cite{Hollik, chmasscorr}. They differ by different
counter terms to the parameters $M'$, $M$ and $\mu$. Although both
schemes are equivalent in the sense that the observables (masses,
cross sections, branching ratios, etc.) are the same, the meaning
of the parameters $M$, $M'$, $\mu$ extracted are different.
In the following we want to analyze in detail the determination of the
parameters of the chargino/neutralino mass matrices at one--loop
level in the scheme \cite{chmasscorr}. We will point out that at one--loop level the
values of the on--shell parameters $M$ and $\mu$ depend on whether
they are determined from the chargino or neutralino system.
Another interesting issue is how the GUT relation $M'=c\,M$
($c=\frac{5}{3}\tan^2\theta_W$ in SU(5), $c=11 \tan^2\theta_W$ in
AMSB) valid in the $\overline{\rm DR}$ scheme can be tested if the
on--shell values of $M'$ and $M$ are extracted from experiment.
The one--loop corrections will also change the
gaugino and higgsino nature of the
charginos and neutralinos, in particular also of the lightest neutralino.
This is important for the dark matter search \cite{darkmatter1, darkmatter2, darkmatter3}.
The whole analysis is based on a full one--loop calculation within
the MSSM. The corresponding formulae are given in the Appendix.

\section{The chargino--neutralino sector}
In the CP conserving MSSM the chargino mass matrix has at tree--level the form
\begin{equation}
        \tilde{X} = \bmat M & \sqrt{2}\, m_W \sin\beta \\
                 \sqrt{2}\, m_W \cos\beta & \mu \emat \,,
  \label{eq:chi+mat}
\end{equation}
where we take $M$, $\mu$, $m_W$ and $\tan\beta$ as on--shell
parameters. $M$ and $\mu$ are taken real.
With the real rotation matrices $U$ and $V$
\begin{equation}
        U = \bmat \cos\Phi_{L} & \sin\Phi_{L} \\
                 -\sin\Phi_{L} & \cos\Phi_{L} \emat \, , \quad
        V = \bmat \cos\Phi_{R} & \sin\Phi_{R} \\
                 -\sin\Phi_{R} & \cos\Phi_{R} \emat
  \label{eq:chi+rotmat}
\end{equation}
it can be diagonalized,
\begin{equation}
U\,\tilde{X}\,V^T=\mbox{diag}(\varepsilon_1\, \tilde{m}_{\cchp_1},\ \varepsilon_2\,
\tilde{m}_{\cchp_2}) \,,
\label{eq:chi+rot}
\end{equation}
with $\varepsilon_i =\pm 1$ and the tree--level masses $\tilde{m}_{\cchp_1} < \tilde{m}_{\cchp_2}$.
The solutions of eq. (\ref{eq:chi+rot}) are
\begin{equation}
\tan 2\Phi_L=\frac{2\sqrt{2}m_W(M\cos\beta+\mu\,\sin\beta)}{M^2-\mu^2-
          2m_W^2\cos 2\beta} \,, \quad
\tan 2\Phi_R=\frac{2\sqrt{2}m_W(\mu\,\cos\beta+M\sin\beta)}{M^2-\mu^2+
          2m_W^2\cos 2\beta}\,,
\label{treeangle}
\end{equation}
and
\begin{equation}
\tilde{m}_{\cchp_{1,2}}^2 =\frac{1}{2} \left( M^2+\mu^2+2m_W^2\mp\sqrt{(M^2+\mu^2+2m_W^2)^2-4(M\mu-m_W^2\sin 2\beta)^2} \right)
\,.\label{treemasses}
\end{equation}
As shown in \cite{chmasscorr}, the on--shell mass matrix $X$ at one--loop level can be written as a sum of the tree--level
mass matrix $\tilde{X}$ in terms of the on--shell parameters as in (\ref{eq:chi+mat}) and
the ultraviolet finite shifts $\Delta X$
\begin{equation}
X=\tilde{X}+\Delta X \,.
\end{equation}
This implies corrections in the mass eigenvalues, $\Delta m_{\cchp_i}$, and in the
rotation angles of the coupling matrices, $\Delta\Phi_L$ and $\Delta\Phi_R$. \\
In the neutralino sector, we have the symmetric tree--level mass matrix
\begin{equation}
    \tilde{Y} = \left(\begin{array}{cccc}
M' & 0 & - m_Z \sin\theta_W \cos\beta & m_Z \sin\theta_W
\sin\beta\\ 0 & M & m_Z \cos\theta_W \cos\beta & -m_Z \cos\theta_W
\sin\beta\\ - m_Z \sin\theta_W \cos\beta & m_Z \cos\theta_W
\cos\beta & 0 & -\mu\\ m_Z \sin\theta_W \sin\beta & -m_Z
\cos\theta_W \sin\beta & -\mu & 0
\end{array}\right) \,.
 \label{neumat1}
\end{equation}
Using the real matrix $Z$, we can rotate from the gauge eigenstate basis of the neutral gauginos and
higgsinos $\psi_j^0=(-i\lambda',-i\lambda^3,\psi_{H_1}^1,\psi_{H_2}^2)$ to
the mass eigenstate basis of the neutralinos
$\cch_i=Z_{ij}\psi_j$,
\begin{equation}
        Z\, \tilde{Y}\, Z^T = \mbox{diag}(\varepsilon_1\,\tilde{m}_{\tilde\chi_1^0},
 \,\varepsilon_2\,\tilde{m}_{\tilde\chi_2^0}, \,\varepsilon_3\,\tilde{m}_{\tilde\chi_3^0},\,\varepsilon_4\,\tilde{m}_{\tilde\chi_4^0}) \,
 .
 \label{neudiag1}
\end{equation}
Taking the one--loop terms into account
\begin{equation}
Y=\tilde{Y}+\Delta Y \,,
\end{equation}
we again obtain corrections in the masses, $\Delta m_{\cch_i}$,
and in the coupling matrix $Z=\tilde{Z}+\Delta Z$.

\section{Parameter fixing}
\begin{figure}[h!]
 \begin{center}
%  \vspace*{-10mm}
 \hspace{-8mm}
 \mbox{\resizebox{80mm}{!}{\includegraphics{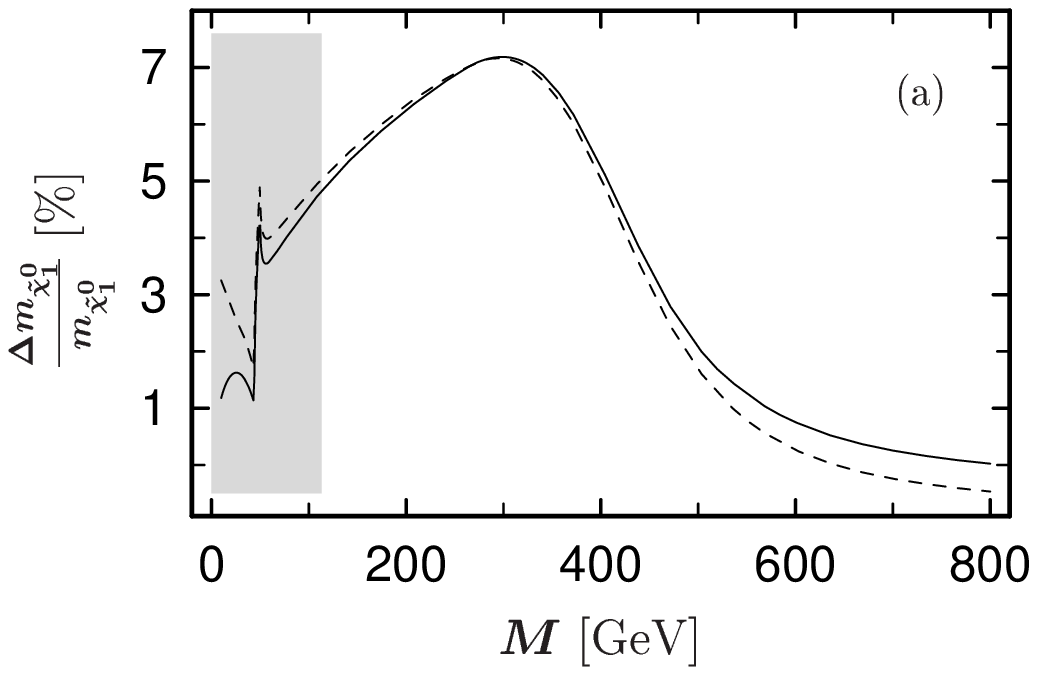}}} % {!}{62mm} == {84mm}{!} !!!!!!
% \hspace{5mm}
 \mbox{\resizebox{80mm}{!}{\includegraphics{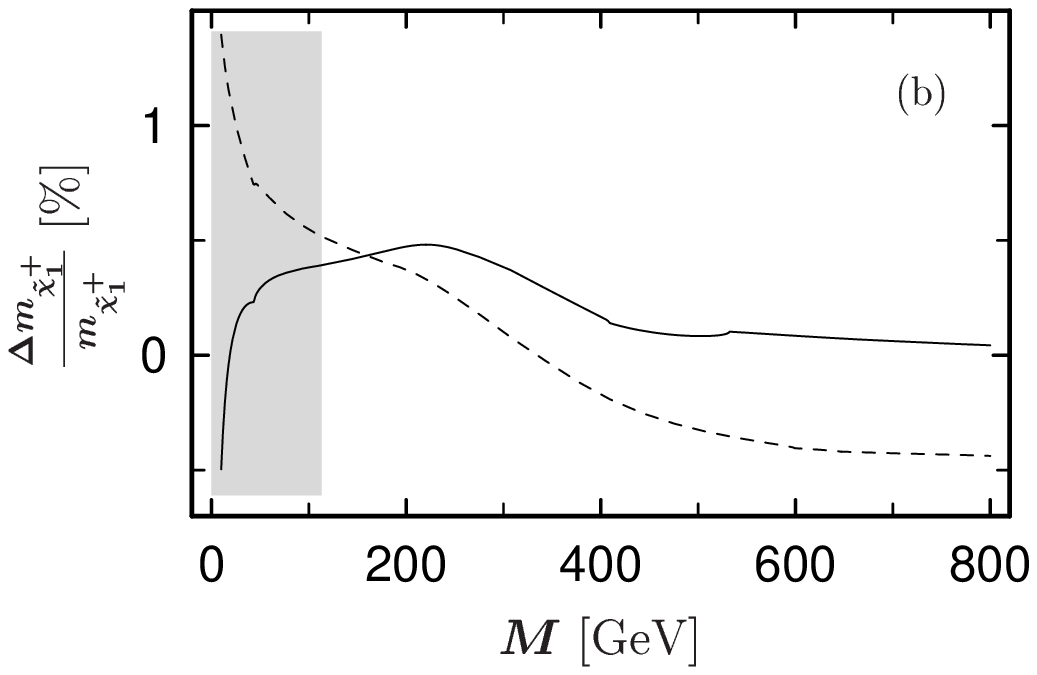}}}
 \hspace{-10mm}
 \vspace{-5mm}
 \caption[fig1]
 {Relative corrections to the $\cch_1$ and $\cchp_1$ masses with gauge
 unification, fixing $M$ and $\mu$ in the chargino (full lines) and
 neutralino (dashed lines) sector.
 The parameters are
 \{$m_{A^0}$, $\tan \b$, $M_{\tilde Q_1}$, $M_{\tilde Q}$, $A$, $\mu$\} =
 \{500, 40, 300, 300, -400, -220\} GeV.
 The grey areas are excluded by the bound $m_{\cchp_1} \geq 100$ GeV.}
 \label{fig:fixing1}
 \end{center}
 \vspace{-7mm}
 \end{figure}

In supersymmetry one has several mass matrices due to the mixing of interaction states. We define the
on--shell mass matrix such that all elements which are non--zero
at tree--level have formally the tree--level form but give the
physical masses and rotation matrices. We always start with a certain
set of on-shell input parameters. For these we need fixing conditions.
All other on-shell entries in the mass matrices can then be calculated.\\
The Standard Model input parameters are the pole masses $m_W = 80.423$~GeV
and $m_Z = 91.1876$~GeV. The Weinberg angle $\theta_W$ is fixed by
$\cos\theta_W = m_W/m_Z$ \cite{sirlin}. The SUSY parameter $\tan\beta$ is fixed
by the condition that there is no transition from the physical CP odd Higgs particle
$A^0$ to the vector boson $Z^0$ \cite{pokorski},
$\textrm{Im}\,\hat{\Pi}_{A^0Z^0}(m_A^2)\,= 0$ which gives
the counter term $\delta\tan{\beta} = {\textrm{Im}\,}{\Pi}_{A^0Z^0}(m_A^2)/(2 m_Z \cos^2\beta)$.
$\hat{\Pi}_{A^0 Z^0}$ is the renormalized self--energy for
the mixing of $A^0$ and $Z^0$.
In this study, the physical input for calculating our input on-shell parameters
$M$, $\mu$, and $M'$ are the two chargino masses and one neutralino
mass. For the other SUSY parameters we use the simplifications
$A_t=A_b=A_\tau=A$ for the trilinear couplings and $M_{\tilde Q_{1,2}}=M_{\tilde
U_{1,2}}=M_{\tilde D_{1,2}}=M_{\tilde L_{1,2}}=M_{\tilde
E_{1,2}}$, $M_{\tilde
Q_{3}}=\frac{10}9M_{\tilde U_{3}}=\frac{10}{11}M_{\tilde
D_{3}}=M_{\tilde L_{3}}=M_{\tilde E_{3}} = M_{\tilde Q}$ for the
soft breaking sfermion mass parameters.\\
In the $\overline{\rm DR}$ scheme (at a scale $Q$) the parameters
$\hat{M}$ and $\hat{\mu}$
are the same in the chargino and neutralino sector. However, the
on--shell parameters $M$ and $\mu$ get different one--loop corrections and thus have different
on--shell values due to different thresholds:
\begin{eqnarray}
\hat{M}(Q)=M+\delta M(Q)=M^{\cchp}+\delta X_{11}=M^{\cch}+\delta
Y_{22} \,,
\\
\hat{\mu}(Q)=\mu+\delta \mu(Q)=\mu^{\cchp}+\delta X_{22}=\mu^{\cch}-\delta Y_{34}
\,.
\end{eqnarray}
$\delta X_{ij}$, $\delta Y_{ij}$ are the counter terms to the
elements $X_{ij}$, $Y_{ij}$ of the chargino and neutralino mass
matrix. The corresponding expressions are given in the Appendix.
The finite difference can be expressed in terms of the chargino
and neutralino mass matrix counter terms
\begin{eqnarray}
\label{dm}
\Delta M&\equiv&M^\cchp-M^\cch =\ \ \ \ \delta Y_{22}-\delta
X_{11}\,, \\ \label{dmu}
\Delta \mu \ &\equiv&\mu^\cchp \ - \ \mu^\cch \ =-(\delta Y_{34}+\delta
X_{22})\,.
\end{eqnarray}
Therefore, we have the freedom to define the input on--shell parameters $M$ and $\mu$ in the chargino sector, i.e. $M\equiv
M^\cchp=X_{11}$, $\mu\equiv\mu^\cchp=X_{22}$, and obtain corrections in the neutralino sector,
or fix $M$ and $\mu$ in the neutralino sector, i.e. $M\equiv
M^\cch=Y_{22}$, $\mu\equiv\mu^\cch=-Y_{34}$, and get corrections in the chargino mass matrix.
For a particular physical situation the elements of the one--loop mass matrices
$X$ and $Y$ (with on--shell parameters plus corrections) are given by the measured neutralino, chargino
masses and other observables, e.g. cross sections.\\
If $M$ and $M'$ are independent parameters it is convenient to use for the
on-shell $M'$ the definition $Y_{11} \equiv M' = \tilde{Y}_{11}$.
If the SU(5) GUT relation, $\hat{M}'=\frac{5}{3}\tan^2\hat{\theta}_W\,\hat{M}$, holds for the $\overline{\rm DR}$
parameters $\hat{M}$ and $\hat{M}'$, we obtain a finite shift for
the on--shell parameters. Thus we can write $Y_{11}\equiv M'=\frac{5}{3}\tan^2\theta_W\,M+\Delta
Y_{11}$, with
 \begin{equation}
  \label{eq:DeltaY11Uni}
 \Delta Y_{11}
 \;=\;\bigg(\frac2{\cos^2{\theta_W}}\,\frac{\delta\sW}\sW \,+\,
 \frac{\delta M}{M}\bigg)\,Y_{11}\;-\;\delta Y_{11}\, .
 \end{equation}
The correction $\Delta Y_{11}$ is due to the same effect and of the same order as $\Delta M$ and $\Delta\mu$, eq.(\ref{dm}) and (\ref{dmu}).
Therefore we include it in our calculations in the cases where gauge unification is explicitly assumed.
Because $M$ depends on the fixing this is also the case
for $\Delta Y_{11}$. Let $\Delta Y_{11}^{\cchp}$
be the correction in the case, where $M$ is fixed in
the chargino sector, and $\Delta Y_{11}^{\cch}$ the case, where
$M$ is fixed in the neutralino sector, it follows
\begin{equation}
\Delta Y_{11}^{\cchp}-\Delta Y_{11}^{\cch}=\frac{5}{3}\tan^2\theta_W\,\Delta
M\,.
\label{dmp}
\end{equation}
In Fig.~\ref{fig:fixing1} the mass corrections
for the lightest neutralino and chargino  assuming gauge unification are shown as a function of $M$. If $M$ and $\mu$
are fixed in the neutralino sector (dashed lines) we have
$Y_{22}\equiv M^{\cch}=M$, $X_{11}\equiv M^{\cchp}=M+\Delta M$,
$Y_{34}\equiv -\mu^\cch=-\mu$, $X_{22}\equiv \mu^\cchp=\mu+\Delta\mu$
and $Y_{11}=\frac{5}{3}\tan^2\theta_WM+\Delta Y_{11}^\cch$. If $M$ and $\mu$ are fixed in the chargino sector (full lines) we
get $Y_{22}\equiv M^{\cch}=M-\Delta M$, $X_{11}\equiv M^{\cchp}=M$,
$Y_{34}\equiv -\mu^\cch=-(\mu-\Delta\mu)$, $X_{22}\equiv \mu^\cchp=\mu$
and $Y_{11}=\frac{5}{3}\tan^2\theta_WM+\Delta Y_{11}^\cchp$.
The differences between the full and the dashed line are due to $\Delta M$ and $\Delta \mu$, eqs. (\ref{dm}), (\ref{dmu}) and (\ref{dmp}).

 \begin{figure}[h!]
 \begin{center}
% \vspace*{-10mm}
 \hspace{-5mm}
 \mbox{\resizebox{84mm}{!}{\includegraphics{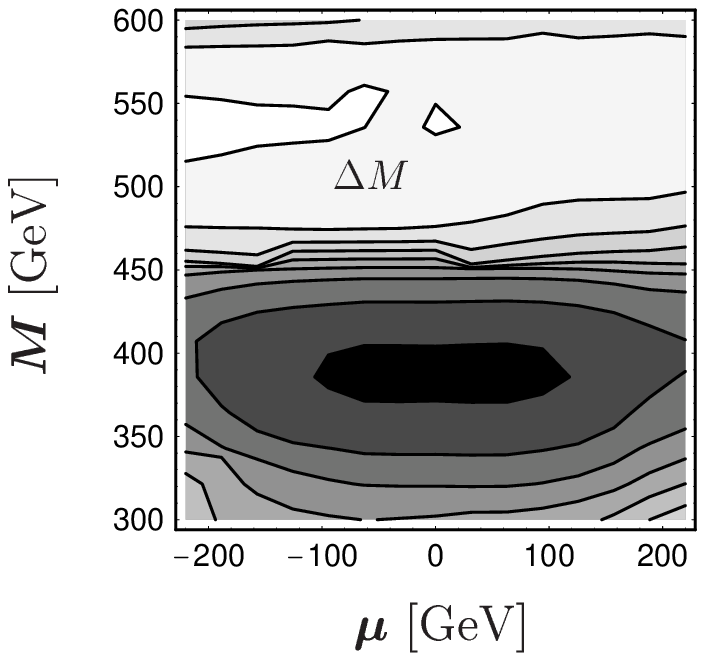}}}
 \hspace{-7mm}
 \mbox{\resizebox{84mm}{!}{\includegraphics{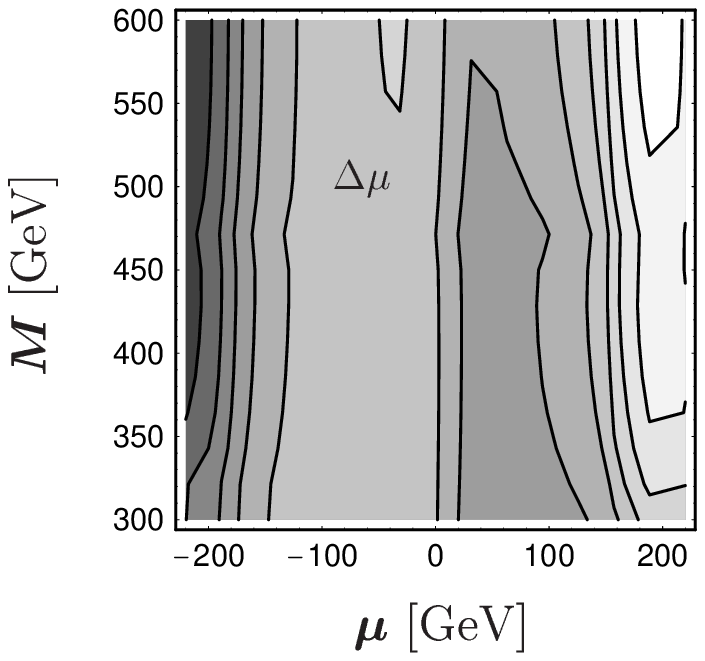}}}
 \hspace{-10mm}
 \vspace{-5mm}
 \caption[fig2]
 {The corrections $\Delta M$ and $\Delta\mu$ as a function of $M$ and $\mu$ (fixed in the chargino sector) with the parameters
 \{$m_{A^0}$, $\tan \b$, $M_{\tilde Q_1}$, $M_{\tilde Q}$, $A$\} =
 \{500, 7, 300, 300, -500\} GeV. $M'$ fulfills the GUT relation.}
 \label{fig:fixing2}
 \end{center}
 \vspace{-7mm}
 \end{figure}

In Fig.~\ref{fig:fixing2} the corrections $\Delta M$ and
$\Delta\mu$ are given as a function of $M$ and $\mu$. For $\Delta
M$ the corrections are in the range of $\Delta M=-0.2$ GeV (white)
and $\Delta M=+0.6$ GeV (black). The corrections $\Delta\mu$ are
between $\Delta\mu=-0.4$ GeV (white)
and $\Delta\mu=+0.5$ GeV (dark grey). The difference between two lines
are $0.1$ GeV.

\section{Coupling corrections}
\begin{figure}[h!]
 \begin{center}
%  \vspace*{-10mm}
 \hspace{-8mm}
 \mbox{\resizebox{80mm}{!}{\includegraphics{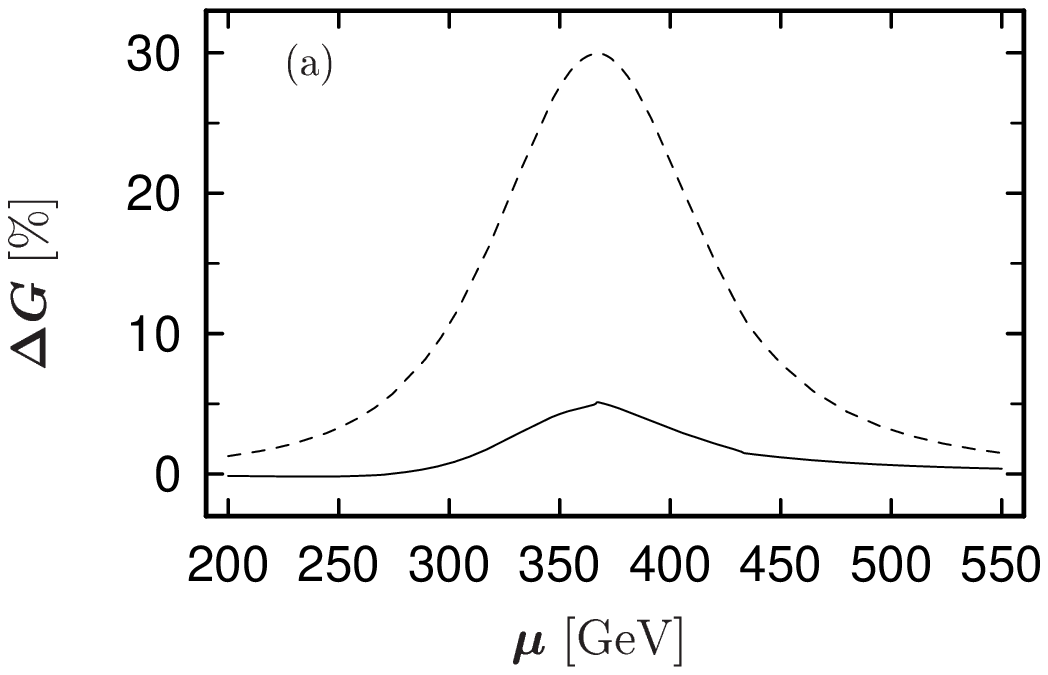}}}
% \hspace{-7mm}
 \mbox{\resizebox{80mm}{!}{\includegraphics{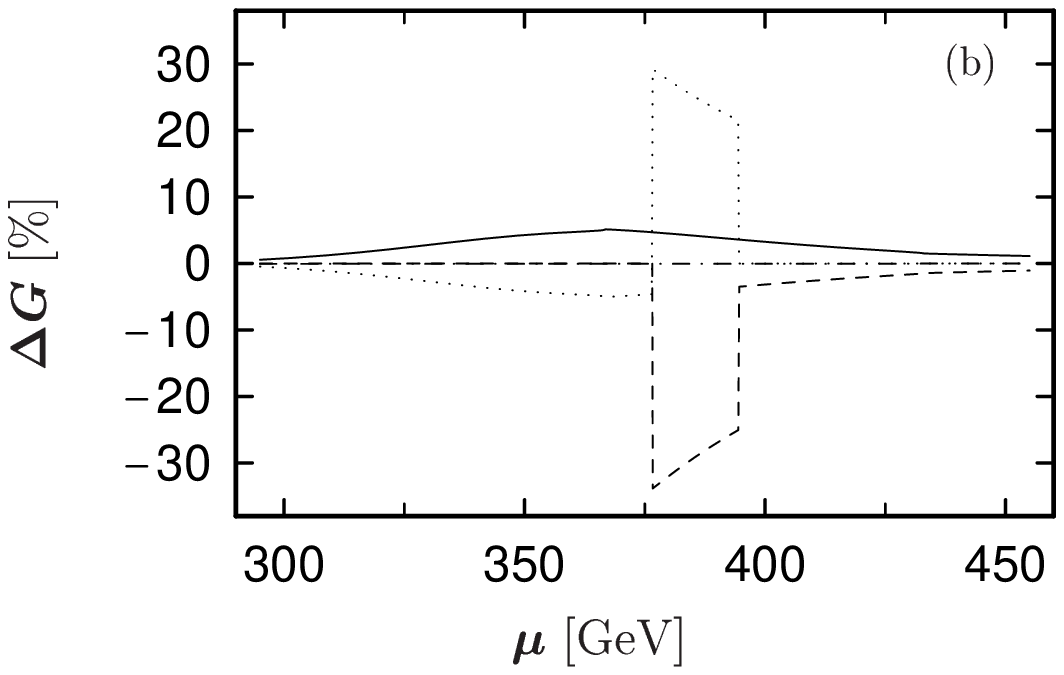}}}
 \hspace{-10mm}
 \vspace{-5mm}
 \caption[fig5]
 {Corrections $\Delta G=\tilde{G} -G$ in the gaugino components as a function of $\mu$, with the parameters
  \{$m_{A^0}$, $\tan\beta$, $M_{\tilde Q_1}$, $M_{\tilde Q}$, $A$, $M$\}
  = \{500, 20, 300, 300, -500, 700\} GeV. In (a) the correction
  of the $\cch_1$ character is presented assuming the SU(5) GUT relation for the on--shell (full line)
  or $\overline{\rm DR}$ (dashed line) parameters  $M$ and $M'$. In (b)
  the corrections for all four neutralinos \{$\cch_1$, $\cch_2$, $\cch_3$, $\cch_4$\} = \{full, dashed, dotted,
  dash--dotted\}
  are given (no gauge unification assumed).
 }
 \label{fig:coupling1}
 \end{center}
 \vspace{-7mm}
 \end{figure}

With the one--loop corrections to the rotation matrices $U$, $V$ and $Z$,
the gaugino and higgsino characters of the individual chargino and
neutralino states change. This can have large effects on decay
widths of processes where these particles are involved \cite{higgs2neut}.
The character of the LSP neutralino plays a key role in dark
matter theories \cite{darkmatter1, darkmatter2, darkmatter3}.
In Fig.~\ref{fig:coupling1} the correction in the gaugino (higgsino) components of
the neutralino $\cch_i$, $G_i = |Z_{i1}|^2+|Z_{i2}|^2$ ($H_i = |Z_{i3}|^2+|Z_{i4}|^2 =
1-G_i$), is presented. In Fig.~\ref{fig:coupling1}a we show that the correction for the lightest
neutralino is in the range of 5\% (full line). In the case of gauge
unification (dashed line) the additional large correction to $Y_{11}$ (approx. +10.8\% at $\mu=370$~GeV)
leads to a change in the gaugino component up to 30\%. In Fig.~\ref{fig:coupling1}b the corrections for all
four neutralinos is given for the same parameter set.
In the range between $\mu=370$ GeV and $\mu=400$ GeV, $\cch_2$ and
$\cch_3$ are nearly mass--degenerated at tree--level. At one--loop
the mass order and -- as a consequence -- the numbering changes.
This is just a small effect in the mass spectrum, but the
interchanging of the gaugino and higgsino components is in the
range of $\pm$30\%.

\section{Parameter analysis}
\begin{figure}[h!]
 \begin{center}
%  \vspace*{-10mm}
 \hspace{-8mm}
 \mbox{\resizebox{80mm}{!}{\includegraphics{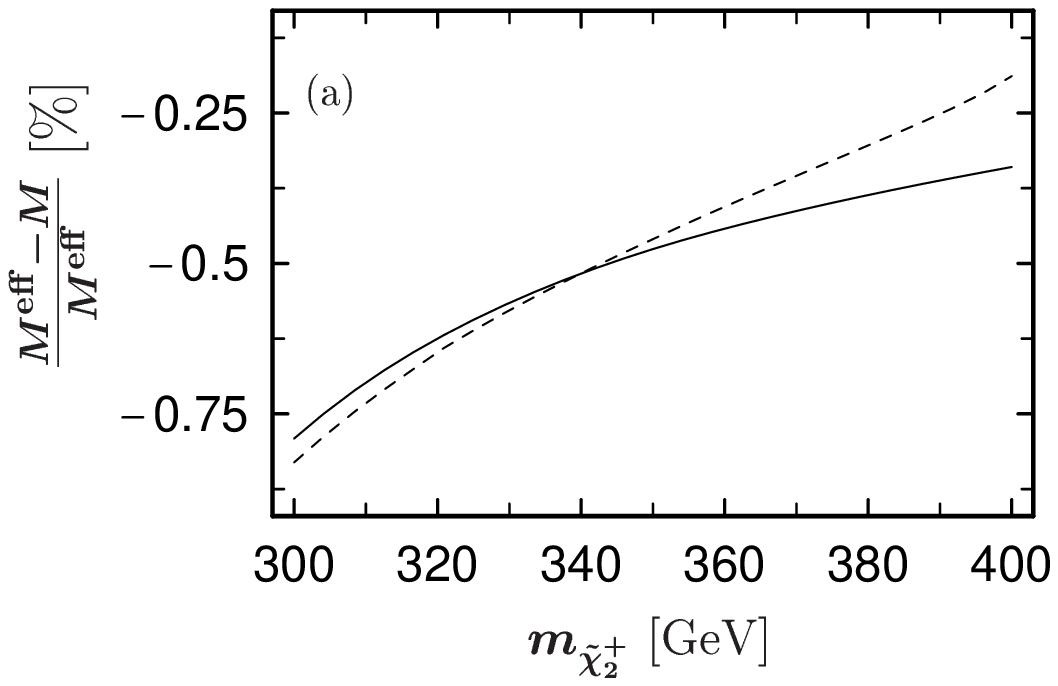}}}
% \hspace{-7mm}
 \mbox{\resizebox{80mm}{!}{\includegraphics{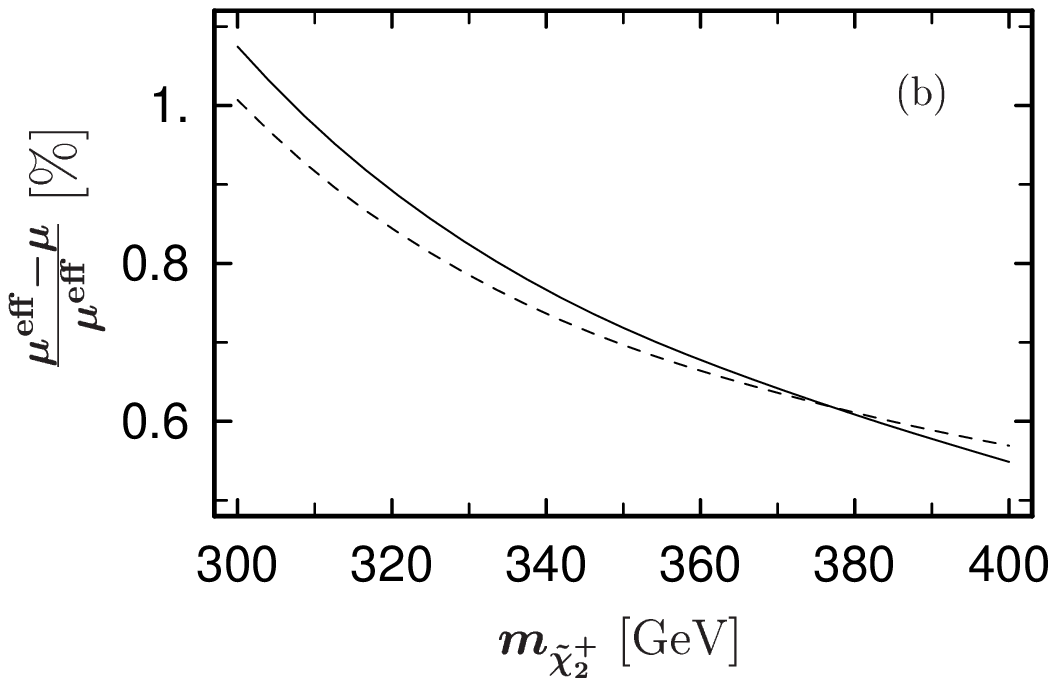}}}
 \hspace{-10mm}
 \vspace{-5mm}
 \caption[fig3]
 {The relative differences between the effective and on--shell $M$ (a) and $\mu$ (b), fixing the on--shell
 parameters in the chargino (full line) and the neutralino (dashed line) sector with
  \{$m_{\cchp_1}$, $m_{\cch_1}$, $\tan \b$, $m_{A^0}$, $M_{\tilde Q_1}$, $M_{\tilde Q}$, $A$\} =
 \{135, 120, 20, 600, 350, 350, 500\} GeV.}
 \label{fig:parameter1}
 \end{center}
 \vspace{-7mm}
 \end{figure}

\begin{figure}[h!]
 \begin{center}
%  \vspace*{-10mm}
 \hspace{-8mm}
 \mbox{\resizebox{80mm}{!}{\includegraphics{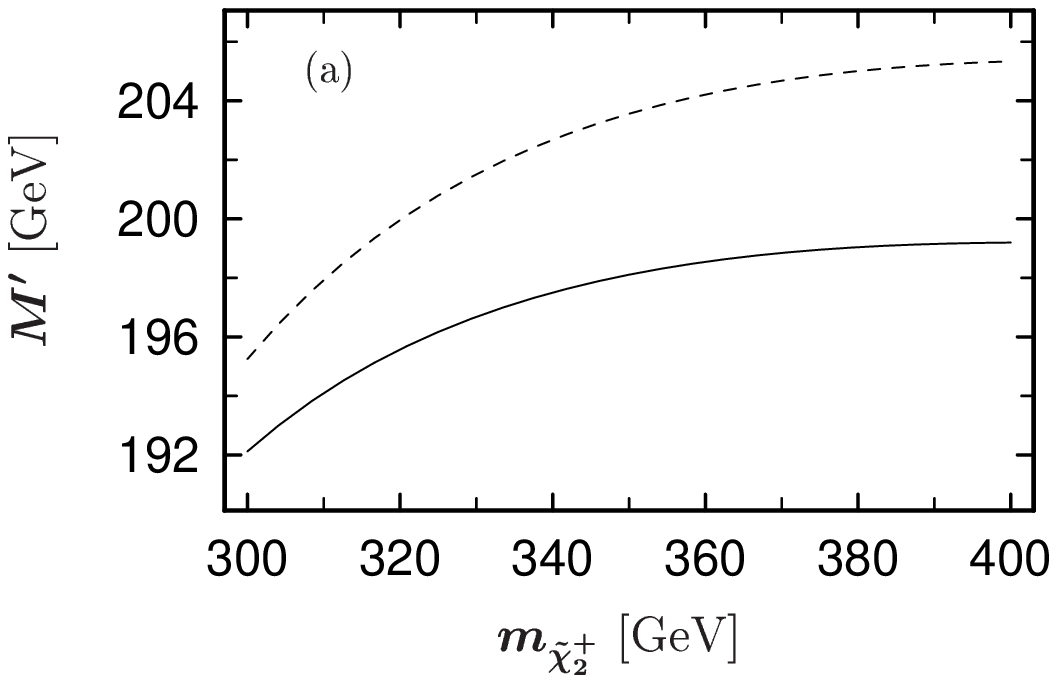}}}
% \hspace{-7mm}
 \mbox{\resizebox{80mm}{!}{\includegraphics{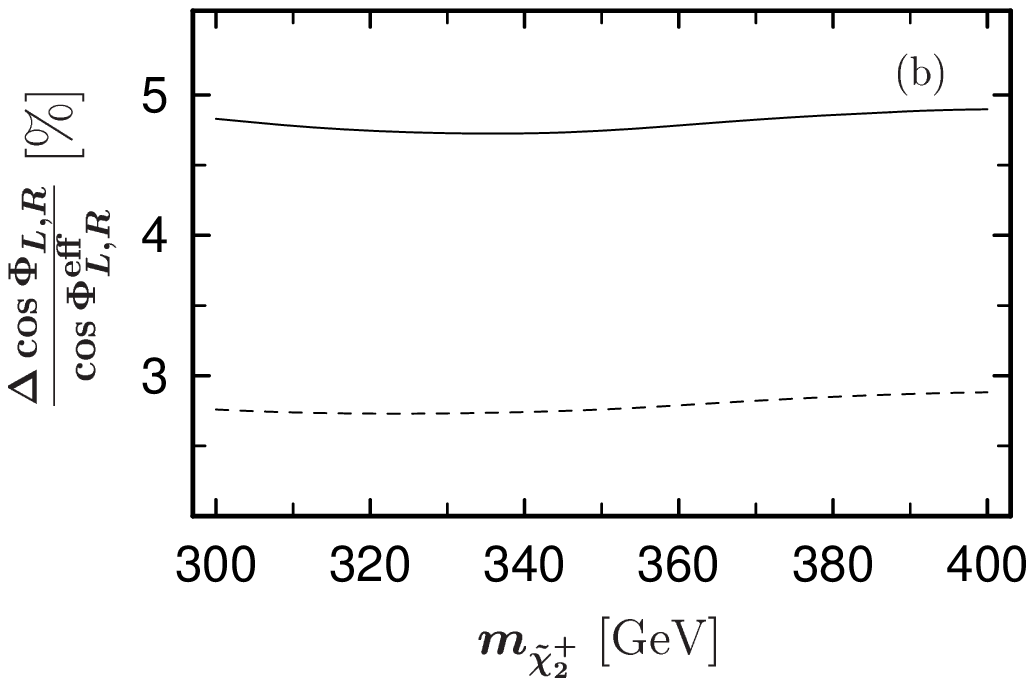}}}
 \hspace{-10mm}
 \vspace{-5mm}
 \caption[fig4]
 {Fig.~(a) shows the effective (dashed line) and on--shell (full line) $M'$. In
  Fig.~(b) there are the relative corrections to $\cos\Phi_L$ (full
  line) and $\cos\Phi_R$ (dashed line). The parameters are the
  same as in Fig.~\ref{fig:parameter1}
 }
 \label{fig:parameter2}
 \end{center}
 \vspace{-7mm}
 \end{figure}

The chargino masses $m_{\cchp_{1,2}}$ and production
cross sections can be measured very precisely at a future $e^+e^-$ linear collider \cite{tesla, lincol}.
From these observables the mixing angles $\cos\Phi_{L,R}$ and by
inverting the relations (\ref{treeangle}) and (\ref{treemasses}) the
fundamental SUSY parameters $M$, $\mu$ and $\tan\beta$ can be
obtained in lowest order \cite{Kalinowski, Kalinowski2}. If a neutralino mass is known, one can also obtain
$M'$ at tree--level.
However, high precision experiments will make it necessary to take into account
one--loop corrections. In the following, we will compare the
tree--level approximations and the full one--loop corrected fundamental SUSY parameters.\\
We use as input the two chargino
masses, the mass of the lightest neutralino and assume the on--shell $\tan\beta$ is known from the Higgs sector \cite{pokorski}.
Calculating the SUSY parameters $M$ and $\mu$ from the tree--level
mass matrices as given in (\ref{eq:chi+mat}) and (\ref{neumat1}) leads to a four--fold ambiguity. For
comparison we choose the same branch as in \cite{Hollik}.
Because we use as input the physical masses this set of tree--level mass
matrices $X^{\rm tree}$, $Y^{\rm tree}$ is different from
$\tilde{X}$ and $\tilde{Y}$, which give the true tree--level mass eigenvalues.
On the other hand, $X^{\rm tree}$ and $Y^{\rm tree}$ are defined
to give the right
physical (on--shell) chargino masses and one neutralino mass.
To calculate the one--loop corrections, values for the other SUSY
parameters ($A$, $M_{\tilde Q}$, $M_{\tilde Q_1}$, $m_{A^0}$) are needed.
The following example, calculated for the same set of parameters as in Fig.~\ref{fig:parameter1}
but $m_{\cchp_2}=350$ GeV, shows the chargino and neutralino mass
matrices in tree--level approximation plus the one--loop
corrections,
%
%\begin{eqnarray}
$$ Y = Y^{\rm tree} + \Delta Y^{\rm tree} = \left(\begin{array}{cccc}
 203.7 & 0 & -2.1 & 42.9\\
 0 & 325.9 & 4.0 & -80.3 \\
 -2.1 & 4.0 & 0 & -147.0 \\
 42.9 & -80.3 & -147.0 & 0  \\
 \end{array}\right)
 \;+\;
 \left(\begin{array}{cccc}
 -5.1 & -0.2 & -0.2 & 1.3\\
 -0.2 & 1.5 & -0.5 &  1.8 \\
 -0.2 & -0.5 & -0.1 & 1.0 \\
 1.3 & 1.8 & 1.0 & 3.7  \\
 \end{array}\right)\, ,
$$
%\nonumber
% \end{eqnarray}

%
%\begin{eqnarray}
$$ X = X^{\rm tree} + \Delta X^{\rm tree} = \left(\begin{array}{cccc}
 325.9 & 113.6 \\
 5.7 & 147.0 \\
 \end{array}\right)
 \;+\;
 \left(\begin{array}{cccc}
 1.6 & -3.1 \\
 -0.9 & -1.1 \\
 \end{array}\right)\, .
$$
%\nonumber
% \end{eqnarray}
Note that both $X^{\rm tree}$ ($Y^{\rm tree}$) and $X$ ($Y$) have
the same physical mass eigenvalues of $\cchp_{1,2}$ and
$\cch_{1}$. We call the parameters used in $X^{\rm tree}$ and
$Y^{\rm tree}$ effective parameters $M^{\rm eff}$, $\mu^{\rm eff}$
and $M'^{\rm eff}$ corresponding to the parameters used in
\cite{Hollik}.
With $Y=\tilde{Y}+\Delta Y=Y^{\rm tree} + \Delta Y^{\rm tree}$
(and the corresponding relation for the chargino mass matrix) the
fundamental on--shell parameters can be determined.
For instance, $M'\equiv Y_{11}=M'^{\rm eff}+\Delta Y_{11}^{\rm
tree}=203.7-5.1$. With $M$ fixed in the chargino system we get
$M\equiv X_{11}=M^{\rm eff}+\Delta X_{11}^{\rm tree}=325.9
+1.6$.\\
In Fig.~\ref{fig:parameter1} and \ref{fig:parameter2}a the
differences between the effective parameters in $Y^{\rm tree}$, $X^{\rm
tree}$ and the properly defined one--loop on--shell parameters
are shown. %($\Delta M=M^{\rm eff}-M$, $\Delta\mu=\mu^{\rm eff}-\mu$)
The effective parameters are obtained applying tree--level relations on the measured masses,
while the on--shell parameters are defined by the elements of the
one--loop corrected mass matrices.
As the effective tree--level and the one--loop corrected chargino mass
matrix have the same eigenvalues, this may imply sizeable corrections
in the rotation angles $\Delta\Phi_{L,R}=\Phi_{L,R}^{\rm eff}-\Phi_{L,R}$. This can be seen
in Fig.~\ref{fig:parameter2}b

\begin{figure}[h!]
 \begin{center}
%  \vspace*{-10mm}
 \hspace{-8mm}
 \mbox{\resizebox{80mm}{!}{\includegraphics{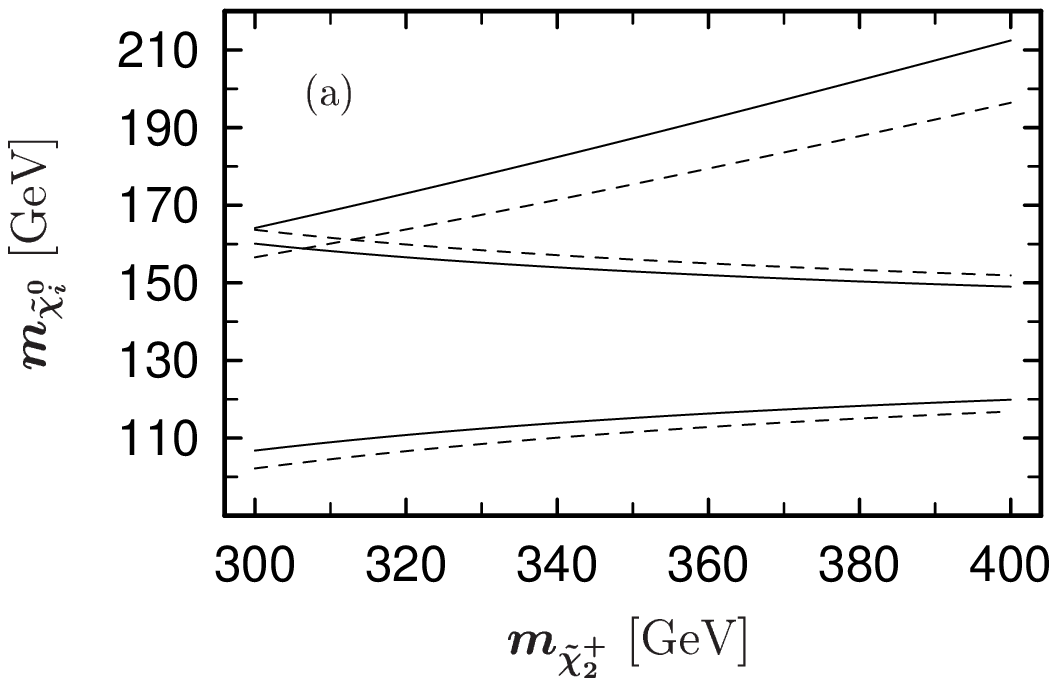}}}
% \hspace{-7mm}
 \mbox{\resizebox{80mm}{!}{\includegraphics{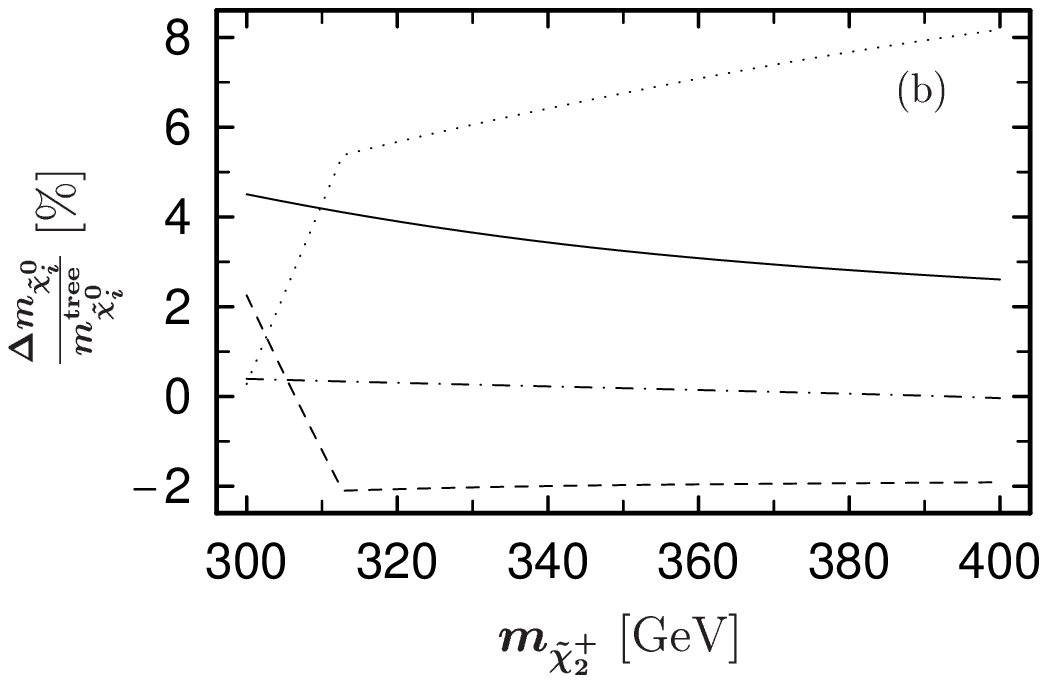}}}
 \hspace{-10mm}
 \vspace{-5mm}
 \caption[fig6]
 {The neutralino mass spectrum (a) at one--loop (full) and tree--level (dashed) and the relative
 corrections (b) for the parameters \{$m_{\cchp_1}$, $\tan \b$, $m_{A^0}$, $M_{\tilde Q_1}$, $M_{\tilde Q}$, $A$\} =
 \{135, 20, 600, 350, 350, 500\} GeV for a SU(5) GUT model. For the mass of the $\cch_4$ (not shown) one has $m_{\cch_4}\simeq
 m^{\rm tree}_{\cch_4}\simeq m_{\cchp_2}$. In (b) the \{full, dashed, dotted,
  dash--dotted\} line corresponds to  \{$\cch_1$, $\cch_2$, $\cch_3$, $\cch_4$\}. 
 }
 \label{fig:gutsu5}
 \end{center}
 \vspace{-7mm}
 \end{figure}

\begin{figure}[h!]
 \begin{center}
%  \vspace*{-10mm}
 \hspace{-8mm}
 \mbox{\resizebox{80mm}{!}{\includegraphics{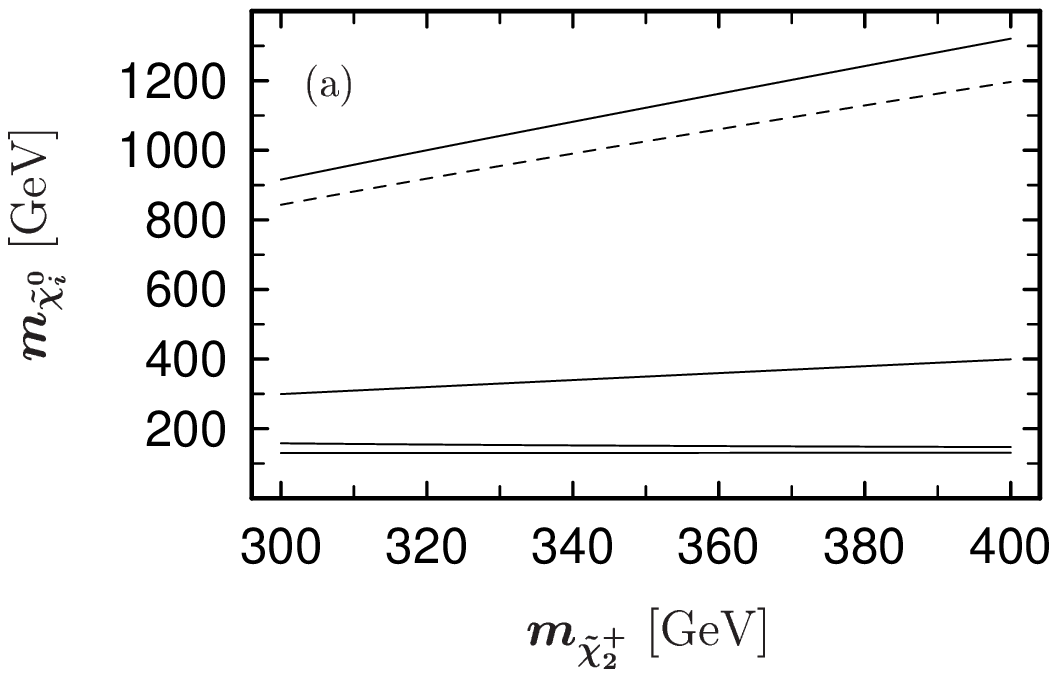}}}
% \hspace{-7mm}
 \mbox{\resizebox{80mm}{!}{\includegraphics{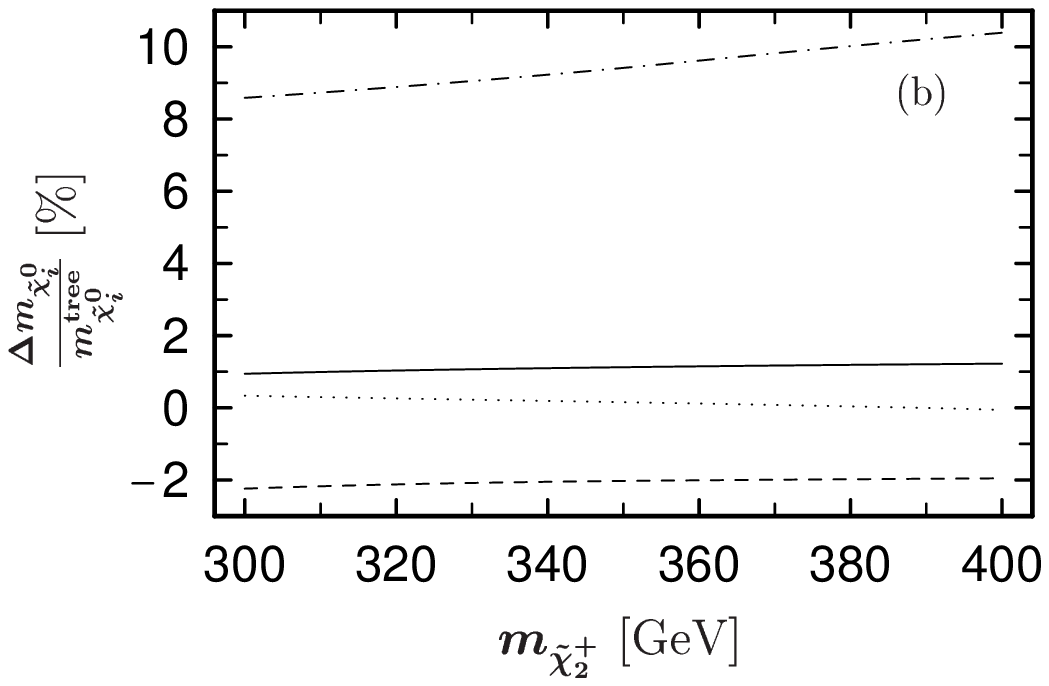}}}
 \hspace{-10mm}
 \vspace{-5mm}
 \caption[fig7]
 {An AMSB model neutralino mass spectrum for the same parameters as in Fig.~\ref{fig:gutsu5}. In (a) only the tree--level
 approximation for the bino--like $m_{\cch_4}$ is shown. In (b) the \{full, dashed, dotted,
  dash--dotted\} line corresponds to  \{$\cch_1$, $\cch_2$, $\cch_3$, $\cch_4$\}. 
 }
 \label{fig:gutamsb}
 \end{center}
 \vspace{-7mm}
 \end{figure}

If the two chargino masses are known from experiment the complete neutralino mass spectrum
can be predicted by assuming the relation $\hat{M}'=c\tan^2\hat{\theta}_W\,\hat{M}$ for the $\overline{\rm DR}$ parameters
or -- in the tree--level approximation -- for the effective parameters.
In Fig.~\ref{fig:gutsu5} and \ref{fig:gutamsb} there are two
different cases shown: For a SU(5) GUT ($c=\frac{5}{3}$) and an AMSB model
($c=11$). We get large corrections for the bino--like neutralino due
to the correction in $Y_{11}$. This is for
$\cch_1$ or $\cch_3$ (depending on $m_{\cchp_2}$) in the SU(5) GUT scenario and 
$\cch_4$ in the AMSB model.

\begin{figure}[h!]
 \begin{center}
 \hspace{-8mm}
 \mbox{\resizebox{87mm}{!}{\includegraphics{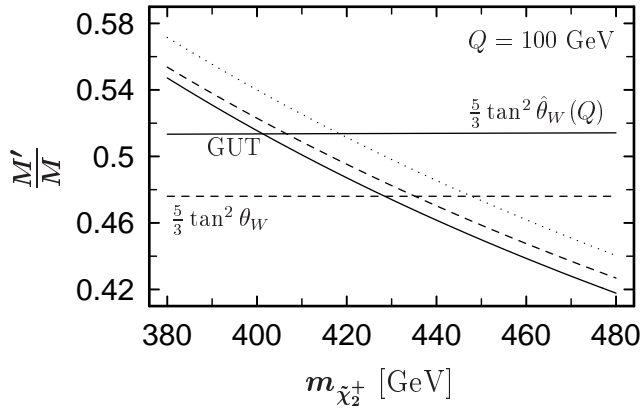}}}
 \hspace{-10mm}
 \vspace{-5mm}
 \caption[fig8]
 {The ratio $\frac{M'}{M\,}$ as a function of $m_{\cchp_2}$. The full, dashed and dotted line corresponds to the $\overline{\rm DR}$, on--shell
 and effective parameters. The input parameters are the same as in
 Fig.~\ref{fig:parameter1}.
 }
 \label{fig:guttest}
 \end{center}
 \vspace{-7mm}
 \end{figure}

The GUT relation can be tested by calculating the
$\overline{\rm DR}$ parameters $\hat{M}(Q) = M + \delta M(Q)$, $\hat{M}'(Q) = M' + \delta M'(Q)$ and
$\tan\hat{\theta}_W(Q) = \tan\theta_W + \delta\tan\theta_W(Q)$ at a scale $Q$. Assuming such a relation for the
on--shell or effective parameters is an inaccurate approximation,
as shown in Fig.~\ref{fig:guttest}.
For the given set of input parameters the ratio
$\frac{\hat{M}'}{\hat{M}\,}$ (full line) fulfills the SU(5) GUT relation
at $m_{\cchp_2} \simeq 402$ GeV.
Using the effective $M^{\rm eff}$, $M'^{\rm eff}$ (dotted line) and the on--shell
$\tan\theta_W$ the calculation leads to $m_{\cchp_2} \simeq 450$
GeV. Even for the on--shell $M$ and $M'$ the GUT point lies
at $m_{\cchp_2} \simeq 437$ GeV.

\section{Conclusions}
We have presented a detailed discussion of the chargino and
neutralino mass parameters at one--loop level. The on--shell
parameters $M$, $\mu$ and $M'$ are properly defined by the on--shell mass
matrix elements.
We have shown that at one--loop level the values $M$ and $\mu$
depend on whether they are determined from the chargino or
neutralino system.
We discussed the difference between the on--shell and the so--called
effective parameters, which are obtained from observables, e.g. on--shell masses, inserted into tree--level
relations.
The corrections to the tree--level mass matrices in terms of the
on--shell and effective parameters are discussed in different scenarios. The numerical analysis based on a complete one--loop
calculation has shown that the corrections to the chargino and
neutralino masses can go up to 10\% and the change in the gaugino
and higgsino components can be in the range of 30\%.
In addition, we have presented how a possible GUT relation for the
parameters $M$ and $M'$ can be tested at one--loop level.

\section*{Acknowledgements}
The authors acknowledge support from EU under the
HPRN-CT-2000-00149 network programme. The work was also supported
by the ''Fonds zur F\"orderung der wissenschaftlichen Forschung'' of Austria,
project no. P13139-PHY.

\begin{appendix}
\section*{Appendix}
\label{sec:app} \setcounter{equation}{0}
\renewcommand{\theequation}{A.\arabic{equation}}
In the following we present the explicit formulas of all
non--(s)fermionic self--energies for the neutralinos, charginos,
$W$ and $Z$ bosons and the $A^0 Z^0$--graphs in the MSSM. The
(s)fermionic part can be found in the appendix of \cite{chmasscorr}.
The two--point functions $A_0$, $B_0$, $B_1$ and $B_{00}$ \cite{thooft} are given
in the convention \cite{denner}.
The neutralino and chargino mass matrix counter terms are
\cite{chmasscorr}:
\begin{eqnarray}
\label{dXY}
\d X_{ij}&=&\frac{1}{2} \sum_{l,n=1}^2 U_{li}V_{nj}\,{\rm Re} \left[ m_{\cchp_l}
\Pi^L_{nl} (m_{\cchp_l}^2) + m_{\cchp_n}
\Pi^R_{ln}(m_{\cchp_n}^2)+\Pi^{S,R}_{nl}(m_{\cchp_l}^2)+\Pi^{S,L}_{ln}(m_{\cchp_n}^2)
\right] \,, \nonumber \\
\d Y_{ij}&=&\frac{1}{2} \sum_{l,n=1}^4 Z_{li}Z_{nj}\,{\rm Re} \left[ m_{\cch_l}
\Pi^L_{nl} (m_{\cch_l}^2) + m_{\cch_n}
\Pi^R_{ln}(m_{\cch_n}^2)+\Pi^{S,R}_{nl}(m_{\cch_l}^2)+\Pi^{S,L}_{ln}(m_{\cch_n}^2)
\right] \nonumber \,,
\\
\end{eqnarray}
with  the convention
\begin{equation}\label{eq:chseconv}
\Pi_{ij}(k^2) =
 \ksla\Big(P_L\Pi^L_{ij}(k^2)+P_R\Pi^R_{ij}(k^2)\Big)
 \,+\,
 P_L \Pi^{S,L}_{ij}(k^2) \,+\,
 P_R \Pi^{S,R}_{ij}(k^2)
 \,.
\end{equation}

\subsection*{Neutralino self--energies}
\begin{eqnarray}
\Pi_{ij}^{H_k^0}(k) \ \ &=& \!\!  \frac{g^2}{(4\pi)^2} \sum_{n=1}^4 \sum_{k=1}^2
\left[ F_{nik}^0 F_{jnk}^0 \left(
m_{\tilde{\chi}_n^0}
B_0(k^2,m_{\tilde{\chi}_n^0}^2,m_{H_k^0}^2)-\ksla
B_1(k^2,m_{\tilde{\chi}_n^0}^2,m_{H_k^0}^2)\right) \right],\hspace{0.8cm} \\
\Pi_{ij}^{A_l^0}(k) \ \ &=& \!\! \frac{-g^2}{(4\pi)^2}\sum_{n=1}^4 \sum_{l=3}^4
\left[ F_{nil}^0 F_{jnl}^0 \left(
m_{\tilde{\chi}_n^0}
B_0(k^2,m_{\tilde{\chi}_n^0}^2,m_{A_l^0}^2)+\ksla
B_1(k^2,m_{\tilde{\chi}_n^0}^2,m_{A_l^0}^2)\right) \right] , \\
\Pi_{ij}^{H_k^+}(k) \hspace{0.1cm} &=& \sum_{n=1}^2 \sum_{k=1}^2
\left[ \left( F_{nik}^R F_{njk}^L + F_{nik}^L F_{njk}^R \right)
m_{\tilde{\chi}_n^+} B_0(k^2,m_{\tilde{\chi}_n^+}^2,m_{H_k^+}^2)-
\right. \nonumber \\
&& \hspace{3.5cm} \left. \left( F_{nik}^R F_{njk}^R + F_{nik}^L F_{njk}^L \right)
\ksla B_1(k^2,m_{\tilde{\chi}_n^+}^2,m_{H_k^+}^2) \right] \,.
\\
\Pi_{ij}^Z(k) \hspace{0.3cm} &=& \frac{2g_Z^2}{(4\pi)^2}
\sum_{n=1}^4 O_{ni}^{''L}O_{jn}^{''L}\left[ 2m_{\tilde{\chi}_n^0}
B_0(k^2,m_{\tilde{\chi}_n^0}^2,m_Z^2)-\ksla
B_1(k^2,m_{\tilde{\chi}_n^0}^2,m_Z^2) \right] \, ,
\\
\Pi_{ij}^W(k) \hspace{0.2cm} &=& \frac{-2g^2}{(4\pi)^2}\sum_{n=1}^2 \left[
2\left(O_{in}^R O_{jn}^L + O_{in}^L O_{jn}^R
\right)m_{\tilde{\chi}_n^+} B_0(k^2,m_{\tilde{\chi}_n^+}^2,m_W^2)+
\right. \nonumber \\ && \hspace{3.5cm} \left.
\left( O_{in}^L O_{jn}^L + O_{in}^R O_{jn}^R \right)\ksla
B_1(k^2,m_{\tilde{\chi}_n^+}^2,m_W^2) \right] \, .
\end{eqnarray}

\subsection*{Chargino self--energies}
\begin{eqnarray}
\Pi_{ij}^{H_k^0}(k) &=& -\frac{g^2}{(4\pi)^2} \sum_{n=1}^2 \sum_{k=1}^2 \left[ \ksla \left(
F_{nik}^+F_{njk}^+ P_R + F_{ink}^+F_{jnk}^+ P_L \right)
B_1(k^2,m_{\tilde \chi_n^+}^2,m_{H_k^0}^2)- \right. \nonumber \\
&& \hspace{20mm} \left. m_{\tilde \chi_n^+} \left(
F_{nik}^+F_{jnk}^+ P_R + F_{ink}^+F_{njk}^+ P_L \right)
B_0(k^2,m_{\tilde \chi_n^+}^2,m_{H_k^0}^2) \right]\,,
\\
\Pi_{ij}^{A_l^0}(k) &=& -\frac{g^2}{(4\pi)^2} \sum_{n=1}^2 \sum_{l=3}^4 \left[ \ksla \left(
F_{nil}^+F_{njl}^+ P_R + F_{inl}^+F_{jnl}^+ P_L \right)
B_1(k^2,m_{\tilde \chi_n^+}^2,m_{A_l^0}^2)+ \right. \nonumber \\
&& \hspace{20mm} \left. m_{\tilde \chi_n^+} \left(
F_{nil}^+F_{jnl}^+ P_R + F_{inl}^+F_{njl}^+ P_L \right)
B_0(k^2,m_{\tilde \chi_n^+}^2,m_{A_l^0}^2) \right] \,,
\\
\Pi_{ij}^{H_k^+}(k) &=& -\frac{g^2}{(4\pi)^2} \sum_{n=1}^4 \sum_{k=1}^2 \left[ \ksla \left(
F_{ink}^L F_{jnk}^L P_R + F_{ink}^R F_{jnk}^R P_L \right)
B_1(k^2,m_{\tilde \chi_n^0}^2,m_{H_k^+}^2)- \right. \nonumber \\
&& \hspace{20mm} \left. m_{\tilde \chi_n^0} \left(
F_{ink}^L F_{jnk}^R P_R + F_{ink}^R F_{jnk}^L P_L \right)
B_0(k^2,m_{\tilde \chi_n^0}^2,m_{H_k^+}^2) \right]\,,
\\
\Pi_{ij}^\gamma (k) &=& -\frac{2e^2}{(4\pi)^2}\delta_{ij}
\left[ \ksla B_1(k^2,m_{\tilde \chi_i^+}^2,0)+2m_{\tilde
\chi_i^+}B_0(k^2,m_{\tilde \chi_i^+}^2,0) \right]\,,
\\
\Pi_{ij}^{Z}(k) &=& -\frac{2g_Z^2}{(4\pi)^2} \sum_{n=1}^2 \left[ \ksla \left(
O_{ni}^{'R}O_{jn}^{'R} P_R + O_{ni}^{'L}O_{jn}^{'L} P_L \right)
B_1(k^2,m_{\tilde \chi_n^+}^2,m_Z^2)+ \right. \nonumber \\
&& \hspace{20mm} \left. 2m_{\tilde \chi_n^+} \left(
O_{ni}^{'R}O_{jn}^{'L} P_R + O_{ni}^{'L}O_{jn}^{'R} P_L \right)
B_0(k^2,m_{\tilde \chi_n^+}^2,m_Z^2) \right]\,,
\\
\Pi_{ij}^{W}(k) &=& -\frac{2g^2}{(4\pi)^2} \sum_{n=1}^4 \left[ \ksla \left(
O_{ni}^{R}O_{nj}^{R} P_R + O_{ni}^{L}O_{nj}^{L} P_L \right)
B_1(k^2,m_{\tilde \chi_n^0}^2,m_W^2)+ \right. \nonumber \\
&& \hspace{20mm} \left. 2m_{\tilde \chi_n^0} \left(
O_{ni}^{R}O_{nj}^{L} P_R + O_{ni}^{L}O_{nj}^{R} P_L \right)
B_0(k^2,m_{\tilde \chi_n^0}^2,m_W^2) \right]\,.
\end{eqnarray}

\subsection*{\boldmath $Z$ self--energies}
\begin{eqnarray}
\Pi_T^{\tilde \chi^0 \tilde \chi^0} (k^2) &=&
-\frac{g_Z^2}{(4\pi)^2}\sum_{i,j=1}^4(O_{ij}^{''L})^2\left[\left(m_{\tilde
\chi^0_i}^2+m_{\tilde \chi^0_j}^2+2m_{\tilde
\chi^0_i}m_{\tilde \chi^0_j}-k^2\right)B_0(k^2,m_{\tilde
\chi^0_i}^2,m_{\tilde \chi^0_j}^2) \right. \nonumber \\
&& \hspace{40mm} \left. +2A_0(m_{\tilde
\chi^0_i}^2)-4B_{00}(k^2,m_{\tilde
\chi^0_i}^2,m_{\tilde \chi^0_j}^2)\right]\, ,
\\
\Pi_T^{\tilde \chi^+ \tilde \chi^+} (k^2) &=&
-\frac{g_Z^2}{(4\pi)^2} \sum_{i,j=1}^2\left[\left( (m_{\tilde
\chi^+_i}^2+m_{\tilde \chi^+_j}^2-k^2)((O_{ij}^{'L})^2+(O_{ij}^{'R})^2)
-4 O_{ij}^{'L} O_{ij}^{'R}m_{\tilde \chi^+_i}m_{\tilde \chi^+_j} \right) \right. \nonumber \\
&& \hspace{-20mm} \left. B_0(k^2,m_{\tilde \chi^+_i}^2,m_{\tilde \chi^+_j}^2)
+ ((O_{ij}^{'L})^2+(O_{ij}^{'R})^2)
\left( 2A_0(m_{\tilde
\chi^+_i}^2)-4B_{00}(k^2,m_{\tilde
\chi^+_i}^2,m_{\tilde \chi^+_j}^2)\right)\right] \,,
\\
\Pi_T^{H_k^0 A_l^0}(k^2) &=& -\frac{g_Z^2}{(4\pi)^2}
\sum_{k,l=1}^2 c_{kl}^2B_{00}(k^2,m_{A_l^0}^2,m_{H_k^0}^2)\,,
\\
\Pi_T^{H_k^+ H_k^+}(k^2) &=& -\frac{g_Z^2}{(4\pi)^2}
(1-2s_W^2)^2 \sum_{k=1}^2
B_{00}(k^2,m_{H_k^+}^2,m_{H_k^+}^2)\,,
\\
\Pi_T^{H_k^0/A_k^0/H_k^+} &=& \frac{g_Z^2}{4(4\pi)^2}\sum_{k=1}^2
\left( A_0(m_{H_k^0}^2) + A_0(m_{A_k^0}^2)+2(1-2s_W^2)^2A_0(m_{H_k^+}^2)
\right) \,,
\\
\Pi_T^{Z H_k^0}(k^2) &=& \frac{g_Z^2}{(4\pi)^2}m_Z^2
\left(
s^2_{\a\b}B_0(k^2,m_{h^0}^2,m_Z^2)+c^2_{\a\b}B_0(k^2,m_{H^0}^2,m_Z^2)
\right)\,, \\
\Pi_T^{W G^+}(k^2) &=& 2\frac{g_Z^2}{(4\pi)^2}m_W^2s_W^4
B_0(k^2,m_{G^+}^2,m_W^2)\,,
\\
\Pi_T^{WW}(k^2) &=& -\frac{g_Z^2}{(4\pi)^2}\left[
10 B_{00}(k^2,m_W^2,m_W^2)+(5k^2+2m_W^2)B_0(k^2,m_W^2,m_W^2) \right. \nonumber \\
&& \hspace{40mm} \left. +2k^2B_1(k^2,m_W^2,m_W^2) +2A_0(m_W^2)\right]\,,
\\
\Pi_T^{W/\omega}(k^2) &=& \frac{3g^2}{8\pi^2}c_W^2 A_0(m_W^2) + 2\frac{g^2}{(4\pi)^2}c_W^2
B_{00}(k^2,m_{\omega^+}^2,m_{\omega^+}^2)\,.
\end{eqnarray}

\subsection*{\boldmath $W$ self--energies}
\begin{eqnarray}
\Pi_T^{\tilde \chi^0 \tilde \chi^+} (k^2) &=& \!\!
\frac{-g^2}{(4\pi)^2}\sum_{i,j=1}^{4,2}\left[\left( (m_{\tilde
\chi^0_i}^2+m_{\tilde \chi^+_j}^2-k^2)((O_{ij}^{L})^2+(O_{ij}^{R})^2)
-4 O_{ij}^{L} O_{ij}^{R}m_{\tilde \chi^0_i}m_{\tilde \chi^+_j}
\right)
 \right.  \\
&& \hspace{-2cm} \left. B_0(k^2,m_{\tilde \chi^0_i}^2,m_{\tilde \chi^+_j}^2)
+ ((O_{ij}^{L})^2+(O_{ij}^{R})^2)
\left(A_0(m_{\tilde
\chi^0_i}^2)+A_0(m_{\tilde \chi^+_j}^2)-4B_{00}(k^2,m_{\tilde
\chi^0_i}^2,m_{\tilde \chi^+_j}^2)\right)\right] \,, \nonumber
\\
\Pi_T^{H H}(k^2) &=& -\frac{g^2}{(4\pi)^2}\left( \sum_{k,l=1}^2 c_{kl}^2
B_{00}(k^2,m_{H_l^+}^2,m_{H_k^0}^2)+\sum_{k=1}^2
B_{00}(k^2,m_{H_k^+}^2,m_{A_k^0}^2) \right) \,,
\\
\Pi_T^{H} &=& \frac{1}{(4\pi)^2}\frac{g^2}{2}\sum_{k=1}^2\left[
\frac{1}{2}A_0(m_{H_k^0}^2)+\frac{1}{2}A_0(m_{A_k^0}^2)+A_0(m_{H_k^+}^2)
\right] \,,
\\
\Pi_T^{H_k^0 W}(k^2)&=&\frac{1}{(4\pi)^2}g^2m_W^2\left(c^2_{\a\b}
B_0(k^2,m_{H^0}^2,m_W^2)+s^2_{\a\b}B_0(k^2,m_{h^0}^2,m_W^2)\right)\, ,
\\
\Pi_T^{Z G^+}(k^2) &=& \frac{1}{(4\pi)^2}m_W^2\left[g_Z^2
s_W^4 B_0(k^2,m_{G^+}^2,m_Z^2)+e^2B_0(k^2,m_{G^+}^2,0)
\right]\,,
\\
\Pi_T^{WZ}(k^2)&=& -\frac{1}{(4\pi)^2}g^2c_W^2\left[
10 B_{00}(k^2,m_Z^2,m_W^2)+(5k^2+2m_Z^2)B_0(k^2,m_Z^2,m_W^2) \right. \nonumber \\
&& \hspace{40mm} \left. +2k^2B_1(k^2,m_Z^2,m_W^2)
+2A_0(m_W^2)\right] \,,
\\
\Pi_T^{W\gamma}(k^2) &=& -\frac{1}{(4\pi)^2}g^2s_W^2\left[
10 B_{00}(k^2,0,m_W^2)+5k^2B_0(k^2,0,m_W^2) \right. \nonumber \\
&& \hspace{40mm} \left. +2k^2B_1(k^2,0,m_W^2)
+2A_0(m_W^2)\right]\, ,
\\
\Pi_T^{Z/\gamma/W} &=& \frac{3g^2}{(4\pi)^2}\left[
c_W^2 A_0(m_Z^2) +s_W^2 A_0(0)+A_0(m_W^2)\right]\, ,
\\
\Pi_T^{\omega}(k^2) &=& 2\frac{g^2}{(4\pi)^2}\left[
c_W^2 B_{00}(k^2,m_{\omega_Z}^2,m_{\omega_+}^2) + s_W^2
B_{00}(k^2,m_{\omega_\gamma}^2,m_{\omega_+}^2) \right]\,.
\end{eqnarray}

\subsection*{\boldmath $A^0 Z^0$ mixing}
\begin{eqnarray}
\Pi_{AZ}^{\tilde \chi^0\tilde \chi^0}(k^2) &=&
\frac{igg_Z}{8\pi^2} \sum_{i,j=1}^4 F_{ji3}^0O_{ij}^{''L}
\left(m_{\tilde \chi^0_i}B_0(k^2,m_{\tilde \chi^0_i}^2,m_{\tilde
\chi^0_j}^2)+(m_{\tilde \chi^0_i}-m_{\tilde\chi^0_j})B_1
(k^2,m_{\tilde \chi^0_i}^2,m_{\tilde \chi^0_j}^2)\right)
\,, \nonumber \\ \\
\Pi_{AZ}^{\tilde \chi^+\tilde \chi^+}(k^2) &=&
\frac{igg_Z}{8\pi^2}\sum_{i,j=1}^2
\left[(F_{ji3}^+O_{ij}^{'L}-F_{ij3}^+O_{ij}^{'R})m_{\tilde
\chi^+_i}B_0(k^2,m_{\tilde \chi^+_i}^2,m_{\tilde
\chi^+_j}^2)+ \right. \nonumber \\
&& \hspace{-20mm} \left. \left( (F_{ji3}^+O_{ij}^{'L}-F_{ij3}^+O_{ij}^{'R})m_{\tilde
\chi^+_i}+(F_{ji3}^+O_{ij}^{'R}-F_{ij3}^+O_{ij}^{'L})m_{\tilde
\chi^+_j} \right)B_1(k^2,m_{\tilde \chi^+_i}^2,m_{\tilde
\chi^+_j}^2)\right]\,,
\\
\Pi_{AZ}^{H}(k^2)&=&\frac{ig_Z^2m_Z}{4(4\pi)^2}\sum_{l,k=1}^2
c_{kl}c_{kl}' \left[ B_0(k^2,m_{A_l^0}^2,m_{H_k^0}^2)+
2B_1(k^2,m_{A_l^0}^2,m_{H_k^0}^2) \right]\,,
\\
\Pi_{AZ}^{ZH}(k^2)&=&\frac{ig_Z^2m_Z}{2(4\pi)^2}
c_{\a\b} s_{\a\b}\sum_{k=1}^2\left[(-1)^k (2B_0+B_1)(k^2,m_Z^2,m_{H_k^0}^2)
\right] \,.
\end{eqnarray}

\subsection*{Couplings}
We used the abbreviations $c_W\equiv\cos\theta_W$,
$s_W\equiv\sin\theta_W$, $g_Z\equiv g / c_W$,
$c_{\a\b}\equiv\cos(\a-\b)$, $s_{\a\b}\equiv\sin(\a-\b)$, with $\alpha$ the mixing angle in the 
\{$h^0$, $H^0$\} system, and for
the Higgs--fields
$H_k^0\equiv\{h^0,H^0\}_k$, $H^\pm_k\equiv\{H^\pm,G^\pm\}_k$,
$A_l^0\equiv\{ A^0, G^0\}_l$.
The coupling matrices are:
\begin{eqnarray} \nonumber
F_{lmk}^0 &=& \hphantom{+}\frac{e_k}{2} \Big[ Z_{l3} Z_{m2} +
Z_{m3} Z_{l2} - \tan\theta_W \left( Z_{l3} Z_{m1} + Z_{m3} Z_{l1}
\right) \Big]
\\[2mm]
&& + \frac{d_k}{2} \Big[ Z_{l4} Z_{m2} + Z_{m4} Z_{l2} -
\tan\theta_W \left( Z_{l4} Z_{m1} + Z_{m4} Z_{l1} \right) \Big] \
= \ F_{mlk}^0
\end{eqnarray}
and
\begin{eqnarray}
F_{ijk}^+ &=& \frac{1}{\sqrt2} \left( e_k V_{i1} U_{j2} - d_k
V_{i2} U_{j1} \right) \,, \\
F_{ilk}^R &=& d_{k+2} \left[ V_{i1} Z_{l4} + \frac{1}{\sqrt2}
(Z_{l2} + Z_{l1} \tan\theta_W) V_{i2} \right] \,,
\\
F_{ilk}^L &=& -e_{k+2} \left[ U_{i1} Z_{l3} - \frac{1}{\sqrt2}
(Z_{l2} + Z_{l1} \tan\theta_W) U_{i2} \right] \,.
\end{eqnarray}
$d_k$ and $e_k$ take the values
\begin{eqnarray}
d_k \ = \ \{-\cos\a, -\sin\a, \cos\b,\sin\b \}_k ,
\quad
e_k \ = \ \{-\sin\a, \cos\a, -\sin\b,\cos\b \}_k \,.
\end{eqnarray}
The other used couplings are
\begin{equation}
O_{ij}^{L} = Z_{i2} V_{j1} - \frac{1}{\sqrt{2}} Z_{i4} V_{j2} \,,
\quad
O_{ij}^{R} = Z_{i2} U_{j1} + \frac{1}{\sqrt{2}} Z_{i3} U_{j2}
\,,
 \label{OLRij}
\end{equation}
\begin{equation}
O_{ij}^{'L} = - V_{i1} V_{j1} - \frac{1}{2} V_{i2} V_{j2} + \delta_{ij}
\sin^2\theta_W = O_{ij}^{'R}(U\leftrightarrow V)\,,
\end{equation}
\begin{equation}
O_{ij}^{''L} = - \frac{1}{2} Z_{i3} Z_{j3} +
\frac{1}{2} Z_{i4} Z_{j4} = - O_{ij}^{''R}\, ,
 \label{Oij''}
\end{equation}
\begin{equation}
c_{kl}:\, c_{11}=c_{22}=c_{\a\b}\,, \ c_{21}=-c_{12}=s_{\a\b} \,,
\end{equation}
\begin{equation}
c_{kl}^\prime = \bmat -\cos 2\beta\sin(\a+\b) &
-\sin 2\beta\sin(\a+\b)\\
\cos 2\beta\cos(\a+\b) &
\sin 2\beta\cos(\a+\b) \emat \, .
 \label{cpkldef}
\end{equation}

\end{appendix}

\end{document}